# WHAT MASSIVE OPEN ONLINE COURSE (MOOC) STAKEHOLDERS CAN LEARN FROM LEARNING ANALYTICS?

*Mohammad Khalil*
*Graz University of Technology, Graz, Austria,*
*mohammad.khalil@tugraz.com*

*Martin Ebner*
*Graz University of Technology, Graz, Austria,*
*martin.ebner@tugraz.com*

## Abstract

Massive Open Online Courses (MOOCs) are the road that led to a revolution and a new era of learning environments. Educational institutions have come under pressure to adopt new models that assure openness in their education distribution. Nonetheless, there is still altercation about the pedagogical approach and the absolute information delivery to the students. On the other side with the use of Learning Analytics, powerful tools become available which mainly aim to enhance learning and improve learners' performance. In this chapter, the development phases of a Learning Analytics prototype and the experiment of integrating it into a MOOC platform, called iMooX will be presented. This chapter explores how MOOC Stakeholders may benefit from Learning Analytics as well as it reports an exploratory analysis of some of the offered courses and demonstrate use cases as a typical evaluation of this prototype in order to discover hidden patterns, overture future proper decisions and to optimize learning with applicable and convenient interventions.

## Key words

MOOCs; Learning Analytics; Evaluation; Visualization; Privacy

## Introduction

Over the past decade, learning has been evolved from its traditional classroom-based forms in a way that is leading to new forms of learning based on technology and distance, moving from a simple idea into a real mainstream. Garrison and Kanuka (2008) showed that the new learning forms using Educational Technology (eLearning) matured into several types of Technology Enhanced Learning, Blended Learning and Online Learning. Different terms for learning through technology have recently come into use, including e-learning, distributed learning, distance learning, web-based learning, tele-learning, and networked learning (Ally, 2004). It is now obvious that the Internet has altered the learning models of educational institutions in schools, academies, and universities. Learning through technology, and specifically online learning, offers flexibility of access anytime and anywhere (Cole, 2000). For example, exchanging information between students and tutors may happen through technology devices such as mobiles and computers. At the moment, students can access learning materials, take quizzes, ask questions, engage with their colleagues and watch learning videos through the Internet. On the other hand, teachers can examine their students' performance through different applications which ease their supervision duties.

Concepts of traditional learning have changed, and the upcoming technologies created new learning environments that did not exist previously. Khalil and Ebner (2015b) listed some of the recent models that are commonly used in Technology Enhanced Learning environments, and these are: "Personal Learning Environments (PLE), Adaptive Hypermedia educational systems, Interactive Learning Environments (ILE), Learning Management Systems (LMS), Learning Content Management Systems (LCMS), Virtual Learning Environments (VLE), Immersive Learning Simulations (ILS), intelligent tutoring systems, mobile learning and MOOCs". Despite the massive quantity of learning contexts, each learning environment is a unique system by itself.

Ever since Siemens and Downes created an open online course in Canada, the MOOCs revolution has been spreading quickly among the fields of online education (McAuley, Stewart, Siemens, & Cormier, 2010). One of the eminent MOOCs movements to have arisen is that which developed after Sebastian Thrun of Stanford University launched a course titled "Introduction to Artificial intelligence" that attracted more than 160,000 learners from different countries around the world (Yuan, Powell, & CETIS, 2013). Since then, MOOCs have reserved a relevant and valuable position in educational practice from various perspectives. For instance, new MOOCs platforms such as Udacity (www.udacity.com) and Khan Academy (www.khanacademy.org) have launched their own learning platform as commercial learning services. By contrast, a non-profit MOOC platform such as edX (www.edx.org) offers courses from prestigious universities, which have proved a major attraction for a larger share of students from all over the world.

With the video lectures, discussion forums and interactivity features, MOOCs are growing massively in numbers. For example, the Open Education Scoreboard[1] already reports more than 2000 MOOCS that are steadily growing. Within this expansion, several issues have evolved into serious dilemmas that affect the different stakeholders in these learning environments. Such issues are the dropout and incompletion rate (Khalil & Ebner, 2014), repetition of learning scenarios, lack of interaction with the instructor (Khalil & Ebner, 2013), difficulties in assessment and stimulating learner motivation (Lackner, Ebner, & Khalil, 2015).

[1] http://www.openeducationeuropa.eu/de/european_scoreboard_moocs(last access August 2015)

Elias (2011) outlined the relatively high-interest potential for data generated by the new distance learning environments and pointed out the birth to what is now termed Learning Analytics. Learning Analytics focuses attention on tools and technologies in order to investigate the data coming from different educational contexts such as online learning environments (Dyckhoff, Zielke, Bültmann, Chatti, & Schroeder, 2012) to enhance teachers' perspectives on how learning is happening. Especially, MOOCs offer certain demands and dilemmas that become accepted as a challenge practice in Learning Analytics approaches (Clow, 2013). An open learning environment such as MOOCs, afford an "exciting opportunity" for the Learning Analytics researchers (Chatti et al., 2014). They play a role as a part of the online learning phenomenon where large quantities of data sets are generated, induced by users who access platforms of the kind. These result in activities which are stored in servers and remain meaningless until they are analyzed. Knox (2014) argued that MOOCs and Learning Analytics seem to be well suited to each other when Learning Analytics promises a technological fix for the problems of educational platforms such as the Massive Open Online Courses issues. The needs for Learning Analytics were thus pressure to overcome MOOCs issues and to unveil hidden information and patterns contained in the large educational data sets. Additionally, the demand for Learning Analytics in MOOCs materialized as an assessment to support future decisions in order to find applicable solutions, optimize learning, and to engage students for a better commitment and success as well as assist courses developers and teachers to improve the power of MOOCs.

According to Clow (2013), and because of the relative newness of MOOCs, research studies that target the combination of Learning Analytics and MOOCs have not yet been extensively carried out and researched. Tabaa and Medouri (2013) mentioned that eminent work in Learning Analytics focuses on Learning Management Systems (LMS) and only a very few have dealt with MOOCs. In addition, the utilization of the vast amounts of data generated in learning environments still limited, and different types of analysis, quantitative and qualitative, are required in order for this to be reflected beneficially on stakeholders (Greller & Drachsler, 2012). The applications of Learning Analytics on MOOCs data sets suffered from a broader research that should support decision makers to enhance learning and its environments (Chatti et al., 2014).

Accordingly, this chapter discusses and describes the experiment of employing Learning Analytics approach on MOOCs platform. It is believed that this approach excels because it was preceded into the area of student performance, based on relations with interactions from online learning environments, focusing in particular on the MOOCs platform. As long as MOOC platforms provide several activities, this prototype is uniqueness in handling data flow and proposes for adequate interventions. Moreover, privacy and ethical issues were considered for a final version release.

## Research Questions

This research study carries out the development phases of a Learning Analytics prototype and its integration into the leading Austrian massive open online courses platform, called iMooX. The authors of this chapter will demonstrate different case studies as a typical evaluation of this prototype. Specifically, the study discusses the interpretation of bulk data as well as spot the light on what MOOCs stakeholders can learn from the traces left by learners. The research study will strongly focus on the Learning Analytics application architecture stages to track learners' activities. In addition, different visualizations and exploratory analysis results will be presented and explained. The study focuses overall on two main research questions:

1. How can the Learning Analytics prototype trace students in a Massive Open Online Course Platforms?
2. What are the patterns and revealed outcomes (evaluation) of applying Learning Analytics in MOOC platforms?

## Research Methodology

This publication concentrates the research work based on a thorough literature study covering the main bifurcation axes: Massive Open Online Courses and Learning Analytics. Specifically, the research study contains two basic directions; each principle depends on the other in order to reach the intended goals. The first direction is the design architecture of the proposed Learning Analytics prototype (Alavi, 1984), and this includes tracing the remnant touches of students, gathering their information, tidying and transforming the data, and storing their information securely in the server database. Furthermore, this step is lengthened by pointing out the procedure of integrating the Learning Analytics prototype into the MOOC platform and the implementation framework. The second part aims to get involved with evaluating the Learning Analytics prototype. Therefore, compound analysis methods and observations were employed on students' data which is collected by the Learning Analytics application. The accumulative generation of users' activities tracks learners and records their actions that yield a noticeable incremental space in database records which is hard to manage. Therefore, an approach of content analysis was used which employs classification and measures the remodeled data (Neuendorf, 2002). Accordingly, the student data has been classified into categories of MOOCs indicators and after that, the data is analyzed and visualized using the R software[2]. Afterwards, several case studies have been examined, and as a consequence to the second research question of discovering hidden information and unveiling patterns in Learning Analytics as in some of the work already carried out such as (Greller & Drachsler, 2012; Khalil & Ebner, 2015a; Taraghi, Ebner, Saranti, & Schön, 2014). This part of the research study inspects quantitative data collection and analysis along with qualitative decisions in order to reveal students' behavior in courses as well as handing insights to MOOC stakeholders.

## Massive Open Online Courses

In the past seven years, Educational Technology witnessed the start of an era for courses of a new type which are massive in terms of student numbers, open for all and are available online. This new type is known as Massive Open Online Courses or more commonly by the abbreviation MOOCs. The term MOOC was first coined in 2008 by David Cormier (Hollands & Tirthali, 2014). The awaited results of MOOCs were different depending on different perspectives. For example, in higher education, institutions were looking forward to improving pedagogical and educational concepts by providing high quality teaching principles and to save costs of university level education. This could happen when an instructor has thousands of students who attend a hypothetical class instead of a physical room which cannot handle groups of more than a hundred of learners. On the other hand, education reformers see a glimmer of hope in the Internet-based models, like MOOCs, which help more students to earn college degrees or certificates at a lower cost to themselves, their families, and the government (Quinton, 2013). The MOOCs objectives thus varied between saving costs and or increasing revenues, improving educational outcomes, extending the reachability as well as accessibility of learning material to everyone (Hollands & Tirthali, 2014) and also providing support for the Open Educational Resources (Ebner, Kopp, Wittke, & Schön, 2014).

[2] http://www.r-project.org

MOOCs provide courses to a diverse type of learners regardless their educational background, gender, age or location. A student from Africa can attend a high quality course provided by Massachusetts Institute of Technology or the Harvard University through their platform (edX) at no cost. All that (s)he needs is an Internet connection. In addition, students are not restricted to one path learning specialization (Johnson, Adams, & Cummins, 2013). For example, a computer animation student has the option to attend an English course or a social science student can enroll in a computer science MOOC class without any limitations.

With the growing number of MOOCs since 2008, it has been noticed that they are split into two main types: cMOOCs which were developed by George Siemens and Stephan Downes based on the philosophy of connectivism, and extended MOOCs or shortly xMOOCs, which are based on classical information transmission (Hollands & Tirthali, 2014). McAuley, Stewart, Siemens, & Cormier (2010) clearly defined MOOCs as "an online course with the option of free and open registration, a publicly shared curriculum, and open-ended outcomes. MOOCs integrate social networking, accessible online resources, and are facilitated by leading practitioners in the field of study. Most significantly, MOOCs build on the engagement of learners who self-organize their participation according to learning goals, prior knowledge and skills, and common interests". The combination of letters in the word "MOOCs" can thus be contextualized as:

- Open: the course needs to be open to everyone without qualifications being required. Accessibility to educational material should be also

assured without limitations. The curriculum, assessment, and the information should be open as well (Rodriguez, 2012).

- Massive: enrollees are much larger than regular classes from hundreds to thousands participants
- Online: No physical attendance is required, and all classes are dealt remotely.

The first real massive open online course by Sebastian Thrun and his colleagues attracted over 160,000 participants from all continents (Yuan, Powell, & CETIS, 2013), and the story of magnetizing more participants continues with the ongoing MOOC providers. As an example, a team from Harvard and Massachusetts Institute of Technology university released their research study on the HarvardX and MITx MOOC platform (edX) in which they examined 1.1 billion logged events of 1.7 million students (Ho et al., 2015). It is a logical development for each MOOC platform to seek influence, achieve popularity and also to attract as many participants as possible (Khalil, Brunner, & Ebner, 2015). Recipients who take part in learning in MOOCs vary in heterogeneity. Some studies and reports show that the vast majority of MOOCs participants are former students who are likely to have access to the higher education (Guo & Reinecke, 2014; Gaebel, 2014; Hollands & Tirthali, 2014). In addition, these studies showed in terms of gender distribution that most students were male and with the greatest proportion of being young learners in MOOCs participants division. Accordingly, with all these growing numbers of participants, the MOOCs audience is becoming heterogeneous and as a consequence of the massive number of enrollees, predicting their categories in advance is becoming an ever more difficult task (Lackner, Ebner, & Khalil, 2015). However, it is summarized that each MOOC depends on (i) learners, and those who register in a MOOC platform and then enroll in one of the courses. (ii) Instructors and those who appear in video lectures, explain the materials to the learners and give assignments. (iii) Context and this includes topics, videos, documents…etc.

The pedagogical approach in MOOCs mainly consists of learning and teaching exchange with the combination of watching videos, downloading course materials, attending quizzes, completing assignments as well as getting in touch in the social discussion forums between the learners themselves and the learners with the course's instructor(s). Taking a deeper look at the pedagogical approaches of MOOCs, Anderson & Dron (2011) explained that distance learning pedagogical models are classified to: connectivism, cognitive-behaviorist and social-constructivist. Rodriguez (2012) postulated that cMOOCs belong to the connectivism which depends on building networks of information, and xMOOCs belong to the cognitive-behaviorist model where guided learning and providing feedback are acquired. On the other hand, Stacey (2014) argued that MOOCs pedagogy is boring and not interactive, unless the online pedagogies are open, connections between the elements of MOOCs which are learners, instructors and context are open on the web, and online learning happens when students are involved in blogs, discussion forums and group assignments. Whilst Yuan, Powell, & Cetis (2013) added that peer assessment techniques and exploiting peer support can revolutionize emergence of new pedagogical models in the massive open online course approaches.

***iMooX Platform & Pedagogy***

iMooX is an online learning stage and the first Austrian xMOOC platform founded in 2013 as a result of a collaboration between the University of Graz and Graz University of Technology. Since the platform went online in February 2014, iMooX has enthralled over 5000 users from different participants target groups. The main idea behind the platform was to introduce explicit Open Educational Resource (OER) courses, keep pace with Open Education and lifelong learning tracks, and to attract a public audience extending from school children to elderly people, or to academic degree holders (Fischer, Dreisiebner, Franken, Ebner, Kopp, & Köhler, 2014). A recent study done in 2015 based on three courses, revealed some demographic information about iMooX (Neuböck, Kopp, Ebner, 2015). The research study showed that 65% of learners were male, 44% were aged between (20-34) years, and 25% were over 50 years old. On the other hand, the educational level status showed that most participants already had an academic degree, whereas less than 10% of students had no graduation or completed a primary school education.

The pedagogical approach of iMooX consists of offering courses to students on a weekly basis. One or more video presented each week in diverse styles (see Figure 1). In addition, documents, interactive learning objects, reference to topics in forums and articles on the web are also offered. Usually, the duration of each course does not exceed more than an eight weeks period with a convenient workload.

xxxxxxxxxxxxxxxxxxxxxxxxxxxxxxxxxxx xxxxxxxxxxxxxxxxxxxxxxxxxxxxxx
xxxxxxxxxxxxxxxxxxxxxxxxxxxxxxxxxxx xxxxxxxxxxxxxxxxxxxxxxxxxxxxxx

*Figure 1.* Videos are presented in diverse styles. Left: personal presentation. Right: Experiment presentation.

The design of the platform endeavor to the cognitive-behaviorist pedagogy theme concepts of Gagne (1965):

- Acquiring the learners' attention and this is done by providing them the correct steps of gaining the learning theory through the online education system.
- Listing the objectives and learning goals of each online course.
- Demonstrate the stimulus by presenting active online learning videos
- Giving feedback through discussion forums and regular emails
- Assessing performance and this is done through computerized assessment of the exams.
- Providing guidance, and this usually depends on learners themselves where self-learning is imperative due to the online learning environment conditions.

Furthermore, the platform also supports social-constructivist pedagogy. It proposes social discussion forums where learners get in touch with instructors as well as information exchange taking place between the students themselves (Khalil & Ebner, 2013).

German is the primary communication language of all courses provided. The online courses are presented on a weekly basis and varied in topics between Science, Technology, Engineering, Mathematics (STEM) as well as history and human sciences. Every week of each course consists of short videos and multiple choice quizzes. The quiz system is fairly different in iMooX platform, in which each student has the option to do five attempts per quiz and the system automatically picks the highest grade. There were two main reasons behind this; from the psychological point of view, the student is less stressed and behaves in a more comfortable manner, whilst researchers can study the participant's learning behavior based on the number of attempts made by the student (Khalil & Ebner, 2015b). iMooX platform offers certificates to participants completely for free, it is only required that students have to successfully finish the quizzes and fill out an evaluation form at the end of each course in which they assess their own experience with the enrolled MOOC.

## Learning Analytics in iMooX

### *Background*

The area of Learning Analytics has developed enormously since the first International Conference on Learning Analytics and Learning in 2011. The emergence of analytics in learning was a reaction to the growing needs of discovering patterns about learners and the needed advice in learning (Siemens, 2010). The proliferation of the Internet and technology and the abundance of data about learners were the major factors that drove the noticeable expansion of Learning Analytics in Educational Technology aspects (Khalil & Ebner, 2015b). A plethora of definitions was used to describe the concept of Learning Analytics before the official one was adapted by the Society for Learning Analytics Research (SoLAR). Siemens (2010) defined it as “the use of intelligent data, learner product data and analysis models to discover information and social connections, and to predict and advise on learning”. In the meanwhile, Elias (2011) described it as the field that is “closely tied” to academic analytics, business intelligence, web analytics, and educational data mining. Learning environments considered as a gold mine of information. Students’ mouse clicks, time spent on questions, their quizzes performance and forums activities are all stored as log files. As a consequence, the fields of educational data mining and analytics seek to use these large amounts of data repositories in order to understand learners and to mutate the practical benefits to them and to the environment where learning happens (Romero & Ventura, 2010). Later on, SoLAR (2011) defined Learning Analytics as “The measurement, collection, analysis and reporting of data about learners and their contexts, for purposes of understanding and optimizing learning and its environment in which it occurs". Ebner, Taraghi, Saranti, & Schön (2015) introduced seven features and the most important directions for smart Learning Analytics.

The purposes of Learning Analytics have been researched in several frameworks (Greller & Drachsler, 2012; Khalil & Ebner, 2015b; Chatti et al., 2014; Greller, Ebner & Schön, 2014), in which the main goals illustrate in creating convenient interventions on learning as well as its environment and the final optimization about learning domain's stakeholders. As a result, the applications of Learning Analytics vary in providing services and tools for the goals of enhancing such learning environments like the MOOC platforms. Clow (2013) pointed out that there is a potential value where Learning Analytics can give a helping hand to learners in a MOOC context. However, integrating Learning Analytics in MOOCs has not been deeply researched and its practices are still limited (Yousef, Chatti, Ahmad, Schroeder, & Wosnitza, 2015; Clow, 2013).

***The iMooX Learning Analytics Prototype***

Boosting learners' motivation and supporting them to improve their learning practice are the intended goals of Learning Analytics in MOOC platforms. A MOOC platform cannot be considered as a real modern educational environment without an analytical approach to examine what is going on. Tracking students' activities in order to reveal hidden patterns thus assures the needs for such a tool to be integrated in the iMooX platform. While browsing courses, learners leave many traces behind them that attract educational data miners and learning analysts. The researchers subsequently mediate and cluster these as useful information for optimizing the learning process. As mentioned above and in reference to the literature study, there are pressing needs for an approach that will help MOOCs stakeholders with their future decisions. The initial main intention was to provide administrators as well as researchers a complete separated tool to examine manners of the students in the MOOC platform. Moreover, the demand by lecturers for a summarization of all activities concerning their learning videos analytics and the attitude of students who attended their courses clearly indicates the urgency of the need for such a tool. Teachers in online learning environments, in which they present their work as videos and assessments, become motivated to evaluate their performance with the involved students in his/her courses (Dyckhoff, Zielke, Bültmann, Chatti, & Schroeder, 2012). Additionally, the massive logs quantity generated by the MOOC platform, required an application to pioneer the data into meaningful information to bring meaningful knowledge for MOOCs stakeholders. Security and ethical principles were considered within the design stages of the Learning Analytics prototype.

The iMooX Learning Analytics prototype is built based on the Learning Analytics framework introduced by Khalil & Ebner (2015b). Accordingly, the same lifecycle was adopted in order to enhance the framework and to apply it successfully with the MOOC platform to glim the educational context of the courses directed toward the benefits of various types of learners. The overall goal of this prototype is to integrate a real analytics tool into a MOOC platform and to render useful decisions based on educational and pedagogical approaches. Currently, the prototype is available for usage by administrators, researchers and decision makers. Instructors can apply for students' results regularly upon request. The iMooX managing institution dedicates diligence to the ethical and security dilemmas and constraints due to the extreme restrictions on the students' privacy regulations in Austria. According to the European Law Directive 95/46/EC[3], there are restrictions on the information disclosure on students until a clear consent or a truly anonymization technique is applied.

*Design Architecture*

The overall design of the Learning Analytics prototype was to propose a tool that provides the MOOC administrators with a proper interpretation of the bulk data that is generated by the learners. It has been taken into account the complexity of log files that the web server produces, which is responsible to pass the students left traces to the Learning Analytics server. A proper processing method with the particularity of being reliable, fast and safe was therefore required for passing the log files in order to present them as readable information. The prototype was developed in virtue of four main stages with a reflective concept of optimizing the learning environment, which is the MOOC platform, and improving the MOOC stakeholders, specifically learners and teachers. Figure 2 shows the main stages of the Learning Analytics prototype design architecture.

xxxxxxxxxxxxxxxxxxxxxxxxxxxxxxxxxxxxxxxxxxxxxxxxxxxx

*Figure 2.* The iMooX Learning Analytics prototype design architecture

The first stage of the design architecture of the Learning Analytics tool is started by generating the data on the learning environment of the MOOC platform. Whenever a user registers an account, enrolls in a course, watches a video or quits a course, this is recorded and results in generating log files. A mass amount of log files leads into what is called "big data". It has been defined as high volume, velocity and variety of unprocessed data that drive

[3] http://eur-lex.europa.eu/LexUriServ/LexUriServ.do?uri=CELEX:31995L0046:en:HTML (last access August 2015)

xxxxxxxxxxxxxxxxxxxxxxxxxxxxxxxxxxxxxxxxxxxxxxxxxxxxxxxxxxxxxxxxxxxxxxx

*Figure 3.* A sample of log files that includes students' activities

into an uneasy job of managing the produced data sets (McAfee, Brynjolfsson, Davenport, Patil, & Barton, 2012). Therefore, a suitable data management and administration has been taken into account with the prototype framework. The next step is settled by the webserver which is responsible for collecting students' information. Gathering users' information is accomplished through tracking users on the MOOC platform. Traces of the students, result in a time-referenced descriptions and accurate content that are gathered for designating features of learners and their interaction activities (Perry & Winne, 2006). In this stage, the system records several interactions such as the logging frequency, the total number of course documents downloads, number of readings in forums, the summation of posts per user, videos interactions, total number of quiz attempts and the quiz scores with the time frame manner of all activities. With all these activities, the stream of information is flowing to the main database pending to be parsed and processed in furtherance of getting visualized to the end user.

Looking forward to the third stage and this is where Learning Analytics operations are performed by parsing the logs and processing them to filter the noisy data, since the data in the log files is unstructured, duplicated and not regularly formatted. The Learning Analytics server is thus programmed to synchronically organize log files and operate semantically to pick up key words that help in detecting the students' activities inside the bulk text file, the log file. These keywords are relevant to what has been coded in the backend to pick the appropriate phrases to distinguish between the students' interactions. The collected data and the process of transforming it should cut the edge into meaningful MOOC indicators that reflects the activities of the users. Figure 3 shows a sample of a raw log file before being processed by the Learning Analytics server.

Finally, the collected and organized data are brought forward to be interpreted and visualized to the end user. In this stage, the Learning Analytics prototype is presented as a User Interface for monitoring purposes and observation. The prototype User Interface is only accessible by researchers and administrators. All the educational data sets collected by the prototype are secured by a Virtual Private Network (VPN) in order to enhance the data protection against unauthorized access. The perception of the visualized results should guide the MOOC stakeholders to (i) benchmark the learning environment and its courses and (ii) improve learner, teacher and administrator progress for meeting the pedagogical practices of iMooX. Learning Analytics should provide powerful tools to support awareness and reflection (Verbert, Duval, Klerkx, Govaerts, & Santos, 2013; Chatti et al., 2014). From the software side, the prototype is intended to show visualizations and to provide noiseless data for researchers, and from the awareness side, reflecting the conclusions of observations on the course developers, learners and teachers is contemplated.

*Implementation Framework*

In this section of the chapter, the implementation framework of the Learning Analytics prototype is presented in figure 4. Simply said, the xxxxxxxxxxxxxxxxxxxxxxxxxxxxxxxxxxxxxxxxxxxxxxxxxxxxxxxxxxxxxxxxxxxxxx
xxxxxxxxxxxxxxxxxxxxxxxxxxxxxxxxxxxxxxxxxxxxxxxxxxxxxxxxxxxxxxxxxxxxxx

*Figure 4.* The iMooX Learning Analytics prototype implementation framework

framework encompasses five steps, it started with the MOOC platform where activities are initiated by the learners. The students' discussions and their interactions with learning videos as well as their progress in quizzes are noted in the log files. These log files are generated by the webserver shown in the figure as the second step. The structure it takes belongs to the Apache HTTP Web Server family[4]. With its convenient Graphical User Interface (GUI), errors management tool and powerful security features, the working environment was pertinent to the desired needs. In the third step, the process proceeds to transfer the log files to the Log Files Management tool. The noisy data is filtered according to the description noted in the previous section, and the flood of logs is organized.

In the fourth step, and this is where the core of the implementation framework resides, the Learning Analytics server parses the incoming log files from the management stage and differentiates between the learners' activities and extracts their timing frames. The server side code is written in Python programming language. Whenever an activity is detected, the information is stored in an intelligent programmed database storage in which researchers have the option to browse it and operate different analysis or educational data mining techniques with high authentication and authorization criteria. This enhances the resilience for additional data

[4] http://httpd.apache.org/

xxxxxxxxxxxxxxxxxxxxxxxxxxxxxxxxxxxxxxxxxxxxxxxxxxxxxxxxxxxxxxx

*Figure 5.* The User Interface of the iMooX Learning Analytics Prototype. (A) User Dashboard. (B) Parameters Dashboard

processes to be added in the future either to the front end user or for research purposes.

Finally, the fifth step is the visualization and the User Interface presentation part of the Learning Analytics prototype. At this stage, the processed data that come from the Learning Analytics server indicating the model learners' MOOC activities are now appropriate to be visualized for the end user. The data are presented in textual format and chart forms, e.g. pie charts, scatter plots, line plots, bar charts…etc. The user can display a full statistics of each user and each course. Figure 5A shows the user dashboard, where administrators can view the student's progress in every course (s)he is enrolled in. The examiner can observe quiz attempts, students' performance as well as the logging frequency in a specified time frame as required. In addition to this, the user interface provides the opportunity to track student activities in downloading documents as well as discussion forums. Nevertheless, for privacy reasons, which will be discussed later, it is not possible to work with the user information in such detail and this is due to the privacy laws and concerns of circumstances that could lead to unwanted ethical breaches such as those that have been discussed in previous studies by (Dyckhoff, Zielke, Bültmann, Chatti, & Schroeder, 2012; Khalil & Ebner, 2015b).

The structure of the User Interface is distinct of supporting an interactive working area by providing a parameter dashboard as shown in figure 5B. The layout of the parameter dashboard tab allows the user to compare two parameters. For instance, relations can be elicited between total posts in the discussion forums and the score of the exams as a meta-statistical case. In addition, the User Interface provides a feature of exporting the results as a document making it applicable to be printed or emailed upon request.

*Privacy and Ethics Consideration*

The collection and processing of student information in Learning Analytics applications could comprise ethical issues in the context of their private data. Eight-dimensional constraints were introduced previously by the authors, and these limit the core advancement of Learning Analytics tools (Khalil and Ebner, 2015). Basically, the issues fall into subjected categories as the following: A) Data accessibility and accuracy. B) Privacy and identification of individuals. C) Disclosure of processed and analyzed information. D) Achieving the Confidentiality, Integrity and Availability (CIA) of data in each Learning Analytics phase. E) Possession and ownership of data. In the Learning Analytics prototype project, the main concerns were to preserve learners' sensitive information. It is a familiar demand that institutions or teachers ask for further information about the analyzed results from the educational datasets.

The requests for a broader information range of the examined datasets may lead to ethical breaches of students' personal information (Greller and Drachsler, 2012). Thus, it is attempted to build an elastic tool that aims to sustain their privacy as well as provide convenient interventions. Additionally, all the examinations and the evaluation phases considered information preservation, while data was kept in a secure server. A research study by Peterson mentioned the needs to keep educational records unveiled to third party businesses or operational functions (Petersen, 2012). In other studies produced in the meantime drew attention to guaranteeing student anonymity in order to avoid embarrassments and exposure of data misuse (Baker, 2013; Slade and Prinsloo, 2013). As a result, a de-identification and anonymization system is under development and will be integrated with the Learning Analytics tool in order to keep the ongoing process of the analysis model while minimizing the risk of harming privacy information disclosure incidents. This system will be built based on the European Data Protection Directive 95/46/EC law of privacy. All student records will be anonymized; on the other hand, each record will have a unique descriptor to guide researchers with their studies. Currently, the tool provides instructors with static documents that show statistics about the course different components while keeping students' Personal Identifiable Information (PII) such as email addresses or photographs confidential.

## Evaluation and Discussion

As a means of entering the evaluation process, the research study method consists of quantitative analysis followed by qualitative decisions in order to extract results out of data. Extrapolating beneficial information from learners' traces is a challenge and requires exploratory analysis rather than hypothesis testing (d'Aquin and Jay, 2013). Visualizations and descriptive statistical models were mainly used to outline different characteristics of the Learning Analytics prototype.

In order to evaluate the prototype, the tool has been implemented in two courses offered by the iMooX platform in 2014. The investigated courses were: "Gratis Online Lernen" and in English "Free Online Learning", abbreviated as (GOL-2014), and "Lernen im Netz" and in English "Learning Online" abbreviated as (LIN-2014). Both of these courses were lectured to students in German. Courses were presented within a rich content that included all the MOOCs interactive components: forums, documents to download, learning videos, and multiple-choice-quizzes. The GOL-2014 course workload was set to be 2 hours/week, starting on 20th October 2014, and ending on 31st December 2014. The lead instructor was a faculty member of Graz University of Technology. Whilst LIN-2014 workload was set to be 5 hours/week, starting on 13th October 2014, and ending on 31st December 2014 and the course's instructor was a faculty member of the University of Graz.

The GOL-2014 was a free course open to anyone and without previous knowledge. The course content was about educating people free through the Internet and giving them tips and tricks of how it can be done. On the other hand, the LIN-2014 was not only a free MOOC, but also a university course counted the students coming from the University of Graz. Its main subject was about giving an overview of trends in learning through mobile, social media and the principles of Open Educational Resources. Every week, a batch of short videos was released for both courses and suggested articles to read were posted on the course's homepage wall. A student must score at least 50% in each GOL-2014 quiz and 75% in LIN-2014 quizzes in order to successfully pass the course, with the ability to repeat a quiz up to five times. The iMooX platform is planned out to consider the highest grade of the five attempts.

The Learning Analytics prototype provides us with a huge amount of information through the MOOC platform. The data were directly collected from both of the examined courses through the process described in figure 2 and figure 4. The examined MOOCs educational data sets include over 100,000 records of events with 1530 students registered. These records contain activities related to discussion forums, documents, videos statistics and quiz scores of each student in each course. In order to make a start on evaluating the collected data from the Learning Analytics prototype, organizing the records and carrying out data transformation and manipulation was required to fulfill the principles of "tidying the data" such as cleaning the messy data sets and mutating them into an easily visualized and structured form (Wickham, 2014). It is worth mentioning that the data manipulation in the evaluation process is different from that in the implementation stage. The data that is processed in the evaluation phase is taken directly from the Learning Analytics server, while the data manipulation in the implementation framework is required for the end user visualization phase where the User Interface layout is presented.

Different use cases will now be presented to point out the potential of Learning Analytics for MOOC stakeholders.

### *Use Case 1: Defining Participants and Dropout*

The previous research studies on the iMooX platform were carried out using surveys and questionnaires (Neuböck, Kopp & Ebner, 2015). However, after the Learning Analytics prototype application was implemented, gathering information about participants in every course offered becomes much more than before. One of the first steps in this evaluation was to generate a general description about the MOOC platform participants. In the first analysis of counting the number of students who were certified and who were registered for both courses, a bar graph was generated to show the differences as shown in figure6.

xxxxxxxxxxxxxxxxxxxxxxxxxxxxxxxxxxxxxxxxxxxxxxxxxxxxxxx

*Figure 6.* General description of the examined courses students

The summary showed that there were 1012 registrants in the GOL-2014 course and 177 students who were handed a certificate, which means a ratio of 17.49% of the total registrants. Whilst the LIN-2014 included 519 registrants and 99 certified students, which make them 19% of the total course registrants.

Categorizing online participants in MOOCs has been a hot topic since 2008. Various studies mentioned categorizing the students based on their engagement and motivation (Kizilcec, Piech, & Schneider, 2013; Hill, 2013; Assan, Li, Ren, & Wen, 2013; Tabaa and Medouri, 2013). By advancing within the same route, and based on the data sets collected from both of the examined courses, the division of participants based on their general activity became as the following:

- Registrants: and those are students who enroll in one of the available courses.
- Active learners: and those are students who at least watch a video, post a thread in the discussion forums or attend a quiz.
- Completers: and those who successfully finish all the quizzes, but do not answer the evaluation form.
- Certified learners: and those who successfully finished all the course quizzes and reviewed their learning experience through the evaluation form.

By gathering the data from the Learning Analytics application, clustering them as above and visualizing the results in figure 7, the analysis showed that both courses have 1531 registrants, 1012 registrants in the GOL-2014 and 519 registrants in LIN-2014. 812 active learners in both courses, 479 active students in GOL-2014 and 333 active students in LIN-2014. 348 completers in which GOL-2104 has 217 students who completed the course and 131 completers in LIN-2014. While there were 276 certified learners in both of the courses, 177 in GOL-2014 and 99 students in LIN-2014.

The evaluation shows a remarkable controversy between registrants and active students. GOL-14 has 47.3% active students, while LIN-2104 has 64.16% active students. The higher completion rate in the LIN-2014 can be explained by those students who belong to the University of Graz who can obtain a total of 4 ECTS if they achieve a pass, which will be added to their university educational records.

xxxxxxxxxxxxxxxxxxxxxxxxxxxxxxxxxxxxxxxxxxxxxxxxxxxxxxxxxxxxxx

*Figure 7.* Number of the examined courses' participants

Talking about the completion rate in MOOCs is a journey in itself. A research study performed by (Jordan, 2013) found that 7.6% is the average completion rate in MOOCs. Furthermore, MOOCs are familiar with high attrition rates and a low motivation environment for learners (Khalil and Ebner, 2014). (Rivard, 2013) stated that a Coursera MOOC called "Bioelectricity" lost 80% of its students before the course actually began, the course finished up with 350 certified students out of 12,700 registrants. Whether the students who gain certificates are to be considered as the perfect students still remains as an ambiguous question. Moreover, it is also still unclear whether completion rates should be referenced to registrants or to the active users. According to Rodriguez (2012), participants in MOOCs can go two different ways: as either lurker or active. Table 1 is thus introduced to show different definitions of dropout rate and their percentages based on different categories of MOOCs participants.

*Table 1*. Different dropout rate definitions based on participant categories in the examined MOOCs

| Course Name | Dropout rate certified to registrants | Dropout rate certified to active stud. | Dropout rate completers to registrants | Dropout rate completers to active stud. | Dropout rate active stud. to registrants |
|---|---|---|---|---|---|
| GOL-2014 | 82.50% | 63.04% | 78.55% | 54.69% | 52.67% |
| LIN-2014 | 80.92% | 70.27% | 74.75% | 60.66% | 35.84% |

Furthermore, the dropout was considered to the students who registered and then fell back. The analysis shows that the students who enrolled (registrants) and became active in the LIN-2014 course were 64.16% with a dropout rate of 35.84% while registrants in the GOL-2014 course dropped by 52.67% to reach 479 active students out of 1012 registrants.

***Use Case 2: Videos Patterns***

In the Learning Analytics prototype, the deployment of the applications such as visualization techniques, data arrangement and the statistical model were deliberated on the level of understanding learners in the MOOC learning environment. Like any other MOOC platform, iMooX depends on video lectures as an elementary approach to deliver the learning content to the students, because of the significant role of the video content in the MOOC platforms. The video lectures are hosted on YouTube; the Learning Analytics prototype mines when a student clicks play, stop or when (s)he watches a video from beginning to end. Figure8 shows a graph line of learners' interaction with four weeks of GOL-2014 learning videos. The turquoise line shows the number of students who pause or skip segment of the videos on a specific second. While the red line shows the number of students who replay the video at a specific second. Figure 8A and figure 8B belong to videos of week1 and week2, it can be noticed that the activity is much higher than the status in videos of week7 and week8 as shown in figure 8C and figure 8D. It is a matter of interest for teachers, researchers and pedagogical experts to examine these portions in order to detect engaging video segments and to inspect students' commitment and behavior through the learning experience.

xxxxxxxxxxxxxxxxxxxxxxxxxxxxxxxxxxxxxxxxxxxxxxxxxxxxxxxxxxxxxxxxxxxxxx

*Figure 8.* GOL-2014 course videos tracking. From top to bottom (A) week1 videos; (B) week2 videos; (C) week7 videos; (D) week8 videos

A case study performed by (Brooks, Thompson, & Greer, 2013) categorized three different types of students on the basis of how they watch videos: engaged rewatcher, regular rewatcher and pauser rewatcher, depending on number of pauses and replays. It has been remarked that most videos activity happens during the first and the last minutes as well as throughout intensive learning content segments. By contrast, video activity decreases through time; it has been noticed that there is a drop in video viewing after the first three weeks in both of the examined courses.

***Use Case 3: Discussion Forums Patterns***

This use case is about analyzing the discussion forums MOOCs indicator, which refers to users' readings and writings. The Learning Analytics prototype mines the discussion forums activities and split them into forum posts and forum reads. The analysis pushed the pedagogical hypothesis, which shows that the more interactive modes for student engagement, the better student learning performance is (Waldrop, 2013). During the course sessions, there were 21,468 reads in the GOL-2014 forums and 9136 reads in the LIN-2014 forums. On the other hand, there were 834 posts in the GOL-2014 and 280 posts in the LIN-2014 course. Several research studies have drawn attention to the significant effect of MOOCs discussion forums for the purposes of providing an enhanced adaptive support to students and groups (Ezen-Can, Boyer, Kellogg, & Booth, 2015). For instance, the head instructor of GOL-2014 commented 116 times (13.90%) of the total number of forum's posts. As a result, the course evaluations, which are submitted by completers, show that this created a friendly atmosphere among students.

Figure 9 demonstrates reading in both of the course forums. On the left, figure 9A, the visualization employs a line graph to show reading activity in the LIN-2014 course. It is obvious that students become less interested in reading in the discussion forums after the first weeks. In figure 9B, the total number of reads reached the highest in the first two days of the GOL-2014 course. The topmost count of reads was on 21st October, which is the first day when videos and content were released. The first week collected 6708 reads, the fourth week gathered around 1700 views and the last week got only 1414 reads.

xxxxxxxxxxxxxxxxxxxxxxxxxxxxxxxxxxxxxxxxxxxxxxxxxxxxxx

*Figure 9.* Students readings in MOOCs discussion forums. From left to right (A) LIN-2014 course forum; (B) GOL-2014 course forum

In summary, it was interesting to find that nearly (50%) of both two courses forums readings' happened by the end of the first two weeks. However, only (10%) was the share of readings in forums in the last two weeks. Moreover, it has also been noticed that reading in both of the forums fell to nearly zero when the courses finished at the end of the year.

xxxxxxxxxxxxxxxxxxxxxxxxxxxx

*Figure 10.* Students posts in MOOCs forums

By the same token, writing in forums did not present a different picture. Figure 10 is a dot plot showing that students wrote more often in the first two weeks and that this period therefore takes the lion's share of the whole number of posts. Each point in the plot represents a student. The maximum number of posts in GOL-2014 was on the first day of the course, with 64 posts. The total number of posts during the course period was 834, with an average of 27.57 posts and a median of 26 posts and there were only 6 posts when the course ended. The LIN-2014 collected 280 posts, with an average of 21.12 posts and median of 5 posts and there were only 2 posts after the course ended.

According to the results of discussion forums analysis, the lead management of iMooX is looking forward to enhancing the social communication between instructors and students as well as providing a solid foundation of peer feedback to attract more students into discussions.

***Use Case 4: Quizzes and Grades***

Almost all MOOCs platforms offer quizzes and exams for students to check their learning understanding. The turn of Learning Analytics illustrates the analysis of students' behavior and their performance. As stated above, iMooX proposes quizzes but in a different form than the traditional method. The students have the opportunity to improve their skills by providing five attempts to pass each quiz. According to a research study by (Ye and Biswas, 2014), lecture watchers and quiz attendees play a major role in defining students' performance in MOOCs. Quiz performance accompanied with downloaded documents and readings in the discussion forums were analyzed. Jiang and his colleagues reported that quiz performance reflects the future proportion of certified registrants (Jiang, Warschauer, Williams, ODowd, & Schenke, 2014). In figure 11, which shows a portion of the GOL-2014 quizzes analysis, a perceptible correlation between students who downloaded documents for the week and their quiz grades for the first attempt was observed. The y-axis is the grade; the x-axis displays file names of each week. Each point represents one student.

xxxxxxxxxxxxxxxxxxxxxxxxxxxxxxxxxxxxxxx

*Figure 11.* Analysis view of GOL-2014 first attempt quizzes compared to files downloads in weeks 1-3.

In the top section, the students who downloaded both of the files scored higher than those who did not download any. The week-one quiz average score for the group that downloaded files (337 users), was (80.7%) and a median of (85%), while the mean for the other group who did not (100 users), was (74.12%) and a median of (71%). In quiz two (417 users), the results were nearly the same, the median was (83%) for both groups. An explanation for this would be that the documents were not crucial enough for the overall grade performance. In week-three quiz (259 users), the difference between both groups was obvious. The mean was (74.2%) for the first group who downloaded the files (187 users), and (59.7%) was the mean for the other group.

In order to maintain student performance, their grades in parallel with their social activity were analyzed. Students of MOOCs, who are engaged in forums, have been found to score better in the exams than who were less active (Cheng, Paré, Collimore, & Joordens, 2011; Coetzee, Fox, Hearst, & Hartmann, 2014). Consequently, a correlation test to compare the students who read in forums and who did ,at least, one quiz (active students) in GOL-2014 course was done as a type of example. Figure 12 is a scatter plot which reveals a relatively weak relation between both factors.

xxxxxxxxxxxxxxxxxxxxxxxxxxxxxxxxxxxxxxxx

*Figure 12.* Relation between reading in forums and students performance in GOL-2014

Y-axis shows a number of readings, but with the use of square root in order to attain an ease of pattern recognition. The x-axis records the average score of all quizzes taken by students. The blue line represents a smooth linear regression line while the gray area around it is the standard error. Students with high performance (Grade > 90) have a "reading in forum" median score of 21 reads. On the other hand, there still students who read more than 20 times, but failed to pass some of the quizzes. Respectively, the standard error area is wider when the grades are less than 60. Nonetheless, it cannot be argued that students who read in forums score better; there are still other factors that influence the overall performance, such as the content of the discussion forums itself, watching the learning videos as well as recommended articles by the tutor.

## Related Work

Various applications were developed to solve the pressing needs of understanding learners and enhancing online learning environments similar to the Learning Analytics prototype. It is realized that the most Learning Analytics applications focused on Learning Management Systems (LMS). However, there have not been many research studies on Learning Analytics practices in MOOCs as already discussed before. For example, Tabaa and Medouri presented a Learning Analytics System for MOOCs (LASyM), which analyzes the huge amount of data generated by MOOCs in order to reveal useful information that can help in building new Platforms and assist in reducing the dropout rate (Tabaa and Medouri, 2013). LASyM lacks consideration of privacy and the extensive analysis can exceed the limits to be reached in the personal student-level data.

Dyckhoff, Zielke, Bültmann, Chatti, and Schroeder (2012) introduced the Learning Analytics Toolkit for Teachers (eLAT) with a simple GUI that requires no knowledge in data mining or analysis techniques. The tool can be used by the teachers to examine their teaching activities and to enhance the general assessments and can be implemented on MOOCs, Moodle and other learning systems. Yousef, Chatti, Ahmad, Schroeder, and Wosnitza (2015) built a Learning Analytics application based on learners' perspective survey to enhance personalization in Learning Analytics practices. Yet, the application has not mentioned if the developers took the personal information of students into consideration. LOCO-Analyst (Learning Object Context Ontology Analyst framework) is another tool that provides teachers with feedback about the students and their performance based on a semantic web (Jovanovic et al., 2008). Additionally, analyzing students patterns in MOOCs were mentioned in different studies recently such as (Ferguson & Clow, 2015) and (Joksimović et al., 2015). The application in this chapter shows promising features to discover and examine the behavior of MOOCs students.

It should be noted that several studies analyzed MOOCs components and the learners' engagement. But finally, it is believed that the Learning Analytics prototype differs from the previous tools and research studies, because it was preceded into the area of student performance, based on relations with indicators from online learning environments, focusing in particular on the MOOCs platform. A de-identification methodology is still under development and will be integrated into the Learning Analytics prototype to anonymize records of students and to protect their identities. Furthermore, there was concentration on the videos' interaction and analysis to answer teachers who adhered to know "Why students skip or replay a video more often at specific seconds?".

## Conclusion

During the past decade, e-learning has evolved into new types of online education that drives the wheel into what is known as MOOCs. This new hype went through different aspects to reach higher education and even school education. With its online platform offering a gold mine of information on students, it has come under the spotlight for researchers in different fields such as educational data mining and Learning Analytics. MOOCs and Learning Analytics seems to be well suited to each other in which learner behaviors appear to suggest greater opportunities of personalization, prediction and discovering hidden patterns in the educational data sets (Knox, 2014).

This chapter discussed further development of a Learning Analytics application that seeks to track and mine students' activities in the lead Austrian MOOC platform, iMooX. During the thorough literature study that was carried out, the limited practices combining both of these fields was noticeable. The main goals were thus to show the experience gained in tracking the traces left by students through their Learning Analytics prototype and to present the results from the assessment of the tool. Stages of the design architecture of the prototype as well as the implementation phases were proposed. Finally, the evaluation process proceeded to analyze two MOOCs offered and to examine case studies in order to review the possibilities for revealing hidden patterns with their potential for showing impressive outcomes that influenced different MOOCs stakeholders.

The future plans for this project are to enhance the de-identification techniques, embellish the visualizations and figures and to improve the feedback that will target the learners themselves.

## Author Information

Mohammad Khalil

Educational Technology

Graz University of Technology, Münzgrabenstraße 35a, Graz 8010, Austria

TEL: +43 660 4522132

Email: mohammad.khalil@tugraz.at

Website: http://mohdkhail.wordpress.com

Biographical Sketch:

Mohammad Khalil is a PhD candidate at the Graz University of Technology. He got his bachelor in computer science, and earned his master degree in information security and digital criminology. Mohammad received a full doctorate scholarship by the European Erasmus Mundus scholarship project. His doctoral studies are generally in Educational Technology and Technology Enhanced Learning. On the other side, his research focuses in Learning Analytics, MOOCs, visualizations, and ethical and privacy issues in the educational analytical approaches.

Martin Ebner

Educational Technology

Graz University of Technology, Münzgrabenstraße 35a, Graz 8010, Austria

Email: martin.ebner@tugraz.at

Website: http://martinebner.at

Biographical Sketch:

Martin Ebner is currently head of the department Social Learning at Graz University of Technology and he is also a senior researcher at the institute of Information Systems and Computer Media. He serves as an international speaker, researcher and national stakeholder in the field of Technology Enhanced Learning. Martin holds an associate professor in media informatics and his work focuses strongly on mobile Learning, Learning Analytics and MOOCs.